\documentclass[letterpaper, 10pt, conference]{ieeeconf}  

\IEEEoverridecommandlockouts                              
\usepackage{cite}
\usepackage{amsmath,amssymb,amsfonts}
\usepackage{amsmath,amssymb,array,graphicx,verbatim,mathrsfs,bm,bbm,mathtools}

\usepackage{dsfont}
\usepackage{booktabs}

\usepackage{algorithm,algorithmic}
\usepackage{mathtools}

\usepackage{amsthm}

\newtheorem{theorem}{Theorem}
\newtheorem{lemma}{Lemma}

\newtheorem{corollary}{Corollary}

\newtheorem{assumption}{Assumption}
\newtheorem{remark}{Remark}
\newtheorem{proposition}{Proposition}
\newtheorem{claim}{Claim}
\usepackage{enumerate}
\usepackage{epstopdf}

\newcommand{\gm}[1]{\mathscr{#1}}

\DeclareMathOperator{\ee}{\mathbb{E}}			
\DeclareMathOperator{\prob}{{P}}			

\newcommand{\rand}{{\textsc{rand}}}

\usepackage{tikz}
\usetikzlibrary{arrows,shapes,calc,positioning}
\tikzstyle{block} = [draw,rectangle, rounded corners, minimum width=1cm, minimum height=0.8cm,text centered, line width=2pt ]
\tikzstyle{arrow} = [thick,->,>=stealth,line width=2pt]

\usepackage{circuitikz}
\usepackage{tikz-cd}
\usepackage{enumerate}
\tikzset{
  shift left/.style ={commutative diagrams/shift left={#1}},
  shift right/.style={commutative diagrams/shift right={#1}}
}
\usetikzlibrary{calc}

\usetikzlibrary{shapes}
\tikzstyle{block} = [draw,rectangle, rounded corners, minimum width=1cm, minimum height=0.8cm,text centered, line width=2pt ]
\tikzstyle{arrow} = [thick,->,>=stealth,line width=2pt]

\renewcommand{\P}{{\mathcal P}}
	
\newcommand{\cteam}{C^{\textsc{team}}}	

\newcommand{\X}{{\mathcal X}}
\newcommand{\U}{{\mathcal U}}

\DeclareMathOperator{\E}{\mathbb{E}}			
\DeclareMathOperator{\Py}{\mathbb{P}}			

\usepackage{algorithmic}
\usepackage{graphicx}
\usepackage{textcomp}

\newcounter{l1}
\newcounter{l2}
\newcounter{l3}
\newcommand{\bdotlist}{\begin{list}{$\bullet$}{}}
\newcommand{\bboxlist}{\begin{list}{$\Box$}{}}
\newcommand{\bbboxlist}{\begin{list}{\raisebox{.005in}{{\tiny $\blacksquare$ \ \ }}}{}}
\newcommand{\bdashlist}{\begin{list}{$-$}{} }
\newcommand{\blist}{\begin{list}{}{} }
\newcommand{\barablist}{\begin{list}{\arabic{l1}}{\usecounter{l1}}}
\newcommand{\balphlist}{\begin{list}{(\alph{l2})}{\usecounter{l2}}}
\newcommand{\bAlphlist}{\begin{list}{\Alph{l2}.}{\usecounter{l2}}}
\newcommand{\bdiamlist}{\begin{list}{$\diamond$}{}}
\newcommand{\bromalist}{\begin{list}{(\roman{l3})}{\usecounter{l3}}}

\def\BibTeX{{\rm B\kern-.05em{\sc i\kern-.025em b}\kern-.08em
    T\kern-.1667em\lower.7ex\hbox{E}\kern-.125emX}}
\begin{document}

\title{\LARGE \bf Optimal Communication and Control Strategies for a Multi-Agent System in the Presence of an Adversary}
\author{Dhruva Kartik, Sagar Sudhakara,  Rahul Jain and Ashutosh Nayyar
\thanks{D. Kartik, S. Sudhakara, R. Jain and A. Nayyar are with the Department of Electrical
Engineering, University of Southern California, Los Angeles, CA 90089
(e-mail: mokhasun@usc.edu; sagarsud@usc.edu; rahul.jain@usc.edu and ashutosn@usc.edu). }
\thanks{This work was supported by National Science Foundation (NSF) grants ECCS 2025732 and ECCS 1750041.}}

\maketitle

\begin{abstract}
We consider a multi-agent system in which a decentralized team of agents controls a stochastic system in the presence of an adversary. Instead of committing to a fixed information sharing protocol, the agents can strategically decide at each time whether to share their private information with each other or not. The agents incur a cost whenever they communicate with each other and the adversary may eavesdrop on their communication. Thus, the agents in the team must effectively coordinate with each other while being robust to the adversary's malicious actions. We model this interaction between the team and the adversary as a stochastic zero-sum game where the team aims to minimize a cost while the adversary aims to maximize it. Under some assumptions on the adversary's capabilities, we characterize a min-max control and communication strategy for the team. We supplement this characterization with several structural results that can make the computation of the min-max strategy more tractable.
\end{abstract}


\section{Introduction}\label{introduction}

In multi-agent systems, the agents may not be able to fully observe the system state and the actions of other agents. A multi-agent system is said to have an \emph{asymmetric} information structure when different agents have access to different information. Each agent must select its actions based only on the limited information available to it.  
Decision-making scenarios with information asymmetry arise in a range of domains such as autonomous driving, power grids, transportation networks, cyber-security of networked computing and communication systems,  and competitive markets and geopolitical interactions (see, for example, \cite{washburn1995two,aumann1995repeated, nayyar2010optimal, nayyar2013decentralized, mahajan2013optimal}). 


Based on the nature of interactions between the agents, multi-agent systems can broadly be classified into three types: (i) teams, (ii) games and (iii) team-games. In teams, all the agents act in a cooperative manner to achieve a shared objective. In  games, each agent has its own objective and is self-interested. In team-games, agents within a team are cooperative but the team as a whole is non-cooperative with respect to other teams.
For agents in the same team, sharing information with each other aids coordination and improves performance. Various information sharing mechanisms \cite{nayyar2013decentralized} arise depending on the underlying communication environment. For instance, if the agents have access to a perfect, costless communication channel, they can share their entire information with each other. On the other hand, if communication is too expensive, the agents may never share their information. In this paper, instead of fixing the information sharing mechanism for  agents in a team, we consider a model in which the  agents can strategically decide whether to share their information with other agents or not. By doing so, the agents in the team can balance the trade-off between the control cost and the communication cost. This joint design of control and communication strategies was considered in \cite{sudhakara2021optimal} and a team-optimal solution was provided using the common information approach \cite{nayyar2013decentralized}.

In some scenarios (e.g. a  battlefield), the team of agents may be susceptible to adversarial attacks. Also, the adversary may have the capability to intercept the communication among the agents. This makes the information sharing mechanism substantially more complicated. While sharing information with teammates may be beneficial for intra-team coordination, it can reveal sensitive information to the adversary. The adversary may exploit this information to inflict severe damage on the system. Such interactions between a team of cooperative agents and an adversary can be modeled as a zero-sum team-game \cite{kartikarxiv}.

In this paper, our focus is on a zero-sum game between a team of two agents and an adversary in which the team aims to minimize the control and communication cost while the adversary aims to maximize it. The system state in this game has three components: a local state for each agent in the team and global state. The adversary controls the global state and each of the agents control their respective local states. We restrict our attention to models in which the agents in the team are more informed than the adversary. Our model allows us to capture several scenarios of interest. For example, the adversary in our model can affect the quality of  and the cost associated with the agents' communication channel and the agents can perfectly or imperfectly encrypt their communication. We analyze a family of such zero-sum team vs. adversary game and provide a characterization of an optimal (min-max) control and communication strategy for the team. This characterization is based on common information belief based min-max dynamic program for team vs. team games discussed in \cite{kartikarxiv}. 

\paragraph{Related Works}

There is a large body of prior work on decision-making in multi-agent systems. In this section, we discuss related works on cooperative teams and team-games. In decentralized stochastic control literature, a variety of information structures (obtained from different information sharing protocols) have been considered \cite{nayyar2010optimal, nayyar2013decentralized, mahajan2013optimal, sudhakara2021optimal}. Another well-studied class of multi-agent teams with asymmetric information is the class of Decentralized Partially Observable Markov Decision Processes (Dec-POMDPs).
Several methods for solving such generic Dec-POMDPs exist in the literature \cite{szer2012maa,seuken2008formal,kumar2015probabilistic,dibangoye2016optimally,rashid2018qmix,hu2019simplified}.

Dynamic games among teams have received some attention over the past few years. Two closely related works are \cite{tang2021dynamic} and \cite{kartikarxiv}. In \cite{tang2021dynamic}, a model of games among teams where players in a team internally share their information with some delay was investigated. The authors of \cite{tang2021dynamic} characterize Team-Nash equilibria under certain existence assumptions. In \cite{kartikarxiv}, a general model of zero-sum games between two teams was considered. For this general model, the authors provide bounds on the upper and lower values of the zero-game. A relatively specialized model was also studied in \cite{kartikarxiv} and for this model, a min-max strategy for one of the teams was characterized in addition to the min-max value. In \cite{bhattacharya2012multi}, the authors formulate and solve a particular malicious intrusion game between two
teams of mobile agents.

The works that are most closely related to our work are \cite{sudhakara2021optimal} and \cite{kartikarxiv}. In \cite{sudhakara2021optimal}, the authors consider a \emph{team} problem in which the agents can strategically decide when to communicate with each other. 
While our model is inspired by the model in \cite{sudhakara2021optimal}, our model is substantially more general and complicated because of the presence of an adversary. In team problems, the agents can use deterministic strategies without loss of optimality, whereas in games, the agents can benefit with randomization. Due to the randomness in agents' strategies and the need to solve a min-max problem as opposed to a simpler minimization problem, different techniques are required for analyzing and solving the team-game.  Our game model is a special case of one of the models studied in \cite{kartikarxiv} and hence, we can use the results in \cite{kartikarxiv} to characterize a min-max strategy for the team. While we borrow some results from \cite{kartikarxiv}, our results on private information reduction in this paper are novel.

\paragraph{Notation}
Random variables are denoted by upper case letters, their realizations by the corresponding lower case letters. In general, subscripts are used as time index while superscripts are used to index decision-making agents. For time indices $t_1\leq t_2$, ${X}_{t_1:t_2}$ is the short hand notation for the variables $({X}_{t_1},{X}_{t_1+1},...,{X}_{t_2})$. Similarly, ${X}^{1:2}$ is the short hand notation for the collection of variables $({X}^1,{X}^2)$.
Operators $\Py(\cdot)$ and $\E[\cdot]$ denote the probability of an event, and the expectation of a random variable respectively.
For random variables/vectors $X$ and $Y$, $\Py(\cdot | {Y}=y)$, $\E[{X}| {Y}=y]$ and $\Py({X} = x \mid {Y} = y)$ are denoted by $\Py(\cdot | y)$, $\E[{X}|y]$ and $\Py(x \mid y)$, respectively. 
For a strategy $g$, we use $\Py^g(\cdot)$ (resp. $\E^g[\cdot]$) to indicate that the probability (resp. expectation) depends on the choice of $g$. For any finite set $\mathcal{A}$, $\Delta\mathcal{A}$ denotes the probability simplex over the set $\mathcal{A}$.
For any two sets $\mathcal{A}$ and $\mathcal{B}$, $\mathcal{F}(\mathcal{A},\mathcal{B})$ denotes the set of all functions from $\mathcal{A}$ to $\mathcal{B}$. We define \textsc{rand} to be mechanism that given (i) a finite set $\mathcal{A}$, (ii) a distribution $d$ over $\mathcal{A}$ and a random variable $K$ uniformly distributed over the interval $(0,1]$, produces a random variable $X \in \mathcal{A}$ with distribution $d$, i.e.,
\begin{align}
X = \rand(\mathcal{A},d,K) \sim d.\label{randmech}
\end{align}
\section{Problem Formulation}\label{sec:problem_formulation}

 Consider a discrete-time  control system with a team of two agents (agent 1 and agent 2) and an adversary. The system comprises of a global state and local states for each agent in the team. Let $X^0_t \in \mathcal{X}^0$ denote the global state and let $X^i_t \in \mathcal{X}^i$ denote the local state of agent $i$. $X_t:=(X^1_t,X^2_t)$ represents the local state of both agents in the team. The initial global state and the initial local states of both agents are independent random variables with state $X^i_1$ having the probability distribution $P_{X^i_1}$, $i=0,1,2$. Each agent perfectly observes its own local state and the global state is perfectly observed by all agents (including the adversary).  Let $U^i_t \in \mathcal{U}^i$ denote the control action of agent $i$ at time $t$. $U_t:=(U^1_t,U^2_t)$ denotes the  control actions of both agents at time $t$. Further, let $U^a_t \in \mathcal{U}^a$ denote the control action of the adversary at time $t$. The global and local states of the system evolve according to 
\begin{align}
	X^0_{t+1}&=k^0_t(X^{0}_{t},U^a_t,W^0_t),\\
    X^i_{t+1}&=k^i_t(X^{0}_{t},X^{i}_{t},U^i_t,W^i_t),~~i= 1,2, \label{dyna}
\end{align}
where $W^i_t \in \mathcal{W}^i$, $i=0,1,2$ is the disturbance in dynamics with  probability distribution $P_{W^i}$. The initial states $X_1^0,X_1^1,X_1^2$ and the disturbances  $\{W^i_t\}_{t=1}^{\infty}$, $i=0,1,2$, are independent random variables.
Note that the next local state of agent $i$ depends on the current local state and control action of agent $i$ \emph{and} the global state. The next global state depends on the current global state and the adversary's action.

In addition to deciding the control actions at each time, the two agents in the team need to decide whether or not to initiate communication at each time. We use the binary variable $M^i_t$ to denote  the communication decision taken by agent $i$. Let $M^{or}_t:=\max(M^1_t, M^2_t)$ and let $Z^{er}_t$ represent the information exchanged between the agents at time $t$. 
In this model when global state $X_t^0 = x$, agents lose packets or fail to communicate with probability $p_e(x)$ even when one (or both) of the agents decides to communicate, i.e. when $M^{or}_t=1$. Here, $p_e:\mathcal{X}^0 \to [0,1]$ maps the global state to a failure probability. Based on the communication model described above we can define variable $Z^{er}_t$ given that $X_t^0 =x$ as:
\begin{equation}
  Z^{er}_t=\begin{cases}
    X^{1,2}_t, ~ w.p. ~1-p_e(x) & \text{if $M^{or}_t=1$}.\\
    \phi, ~~~~~ w.p. ~~~~p_e(x) & \text{if $M^{or}_t=1$}.\\
    \phi, & \text{if $M^{or}_t=0$}.
  \end{cases} \label{erasure_model}
\end{equation}
At time $t^+$, the adversary observes a noisy version $Y_t$ of the variable $Z_t^{er}$ given by
\begin{align}
    \label{advobs}Y_t = l_t(Z_t^{er},M_t,X_t^0, W_t^y),
\end{align}
where $W_t^y$ is the observation noise.

\textbf{Information structure and decision strategies:}
At the beginning of the $t$-th time step, the information available to agent $i$ is given by (i) history of global states and its local states, (ii) its control actions, (iii) communication actions and messages and (iv) adversary's action and observation history:
\begin{align}
    &I^i_{t} = \\
    &\{X_{1:t}^0,X^i_{1:t},U^i_{1:t-1},M^{1,2}_{1:t-1},Z^{er}_{1:t-1}, U_{1:t-1}^a,Y_{1:t-1}\}.\nonumber
\end{align}
Agent $i$ can use this information to make its communication decision at time $t$. We allow the agent to randomize its decision. Thus, agent $i$ first selects a distribution $\delta M^i_{t}$ over $\{0,1\}$ based n its information and then it randomly picks $M^i_t$ according to the chosen distribution:
\begin{align}\label{commact}
    \delta M^i_{t} &= f^i_t(I^i_{t}),\\
    M_t^i&=\rand(\{0,1\},\delta M_t^i,K_t^i)
\end{align}
where $K_t^i, i=1,2, t \geq 1,$ are independent random variables uniformly distributed over the interval $(0,1]$ that are used for randomization (these variables are also independent of initial states and all noises/disturbances). The function $f^i_t$ is referred to as the communication strategy of agent $i$ at time $t$. At this point, the adversary does not take any action.
After the communication decisions are made and the resulting communication (if any) takes place, the information available to  agent $i$ is 
\begin{align}
    I^i_{t^+}&=\{I^i_{t},Z^{er}_t,M^{1,2}_t,Y_t\}.
\end{align}

$I_t^a$ denotes the adversary's information just before  the communication at time $t$    and  $I_{t^+}^a$ denotes the adversary's information after communication at time $t^+$. Our model allows for different scenarios of adversary's information which will be described later.

Agent $i$ and the adversary  choose their control actions based on their post-communication information according to 
\begin{align}\label{controlact}
    \delta U^i_{t}& = g^i_t(I^i_{t^+})\\
    U^i_{t} &= \rand(\mathcal{U}_t^i,\delta U_t^i,K_{t^+}^i),~~i=1,2,a,
\end{align}
where  $K_{t^+}^i, i=1,2, t \geq 1,$ are independent random variables uniformly distributed over the interval $(0,1]$ that are used for randomization (these variables are also independent of all other randomization variables,  initial states and all noises/disturbances). The functions $g^i_t$ and $g_t^a$ are referred to as the control strategy of agent $i$ and the adversary at time $t$.
 The tuples $f^i:=(f^i_1,f^i_2,...,f^i_T)$ and  $g^i:=(g^i_1,g^i_2,...,g^i_T)$
are called the communication and control strategy of agent $i$ respectively.
 The collection $f:=(f^1,f^2)$, $g:=(g^1,g^2)$ of communication and control strategies of both agents are called the communication and control strategy of the team. Similarly, $g^a:=(g^a_1,g^a_2,...,g^a_T)$ is called the control strategy of the adversary.

We can split the information available to the agents into two parts -- common information and private information. Common information at a given time is the information available to all the decision-makers (including the adversary) at the given time. Private information of an agent includes all of its information at the given time except the common information. \begin{enumerate}
\item At the beginning of time step $t$, before the communication decisions are made, the common ($C_t$) and private information ($P_t^i$) is defined as 
\begin{align}
C_t &:= I_t^1 \cap I_t^2 \cap I_t^a\\
P_t^i &:= I_t^i \setminus C_t \quad \forall i \in \{1,2,a\}.
\end{align}
\item After the communication decisions are made and the resulting communication (if any) takes place, the common and private information is defined as 
\begin{align}
C_{t^+} &:= I_{t^+}^1 \cap I_{t^+}^2 \cap I_{t^+}^a\\
P_{t^+}^i &:= I_{t^+}^i \setminus C_{t^+} \quad \forall i \in \{1,2,a\}.
\end{align}
\end{enumerate} 
\begin{assumption}\label{infoassump}
We assume that the following conditions are satisfied:
\begin{enumerate}
    \item \emph{Monotonicity:} The adversary's information grows with time. Thus, $I_t^a \subseteq I_{t^+}^a \subseteq I_{t+1}^a$ for every $t$. 
    \item \emph{Nestedness:} The adversary's information is common information and each agent in the team has access to adversary's information, i.e.,
    \begin{align}
         C_t=I_{t}^a &\subseteq I_{t}^1 \cap I_{t}^2 =: \cteam_t \nonumber\\
        C_{t^+}=I_{t^+}^a &\subseteq I_{t^+}^1 \cap I_{t^+}^2 =: \cteam_{t^+} .\nonumber
    \end{align}
    Therefore, $P_t^a = P_{t^+}^a = \varnothing$.
    \item \emph{Common Information Evolution:}
(i)  Let ${Z}_{t^+} \doteq {C}_{t^+}\setminus {C}_t$ and ${Z}_{t+1} \doteq {C}_{t+1}\setminus {C}_{t^+}$ be the increments in common information at times $t^+$ and $t+1$, respectively . Thus, ${C}_{t^+} = \{{C}_t,{Z}_{t^+}\}$ and ${C}_{t+1} = \{{C}_{t^+},{Z}_{t+1}\}$. The common information evolves as
\begin{align}
\label{commonevol}{Z}_{t^+} &= \zeta_{t^+}({P}_t^{1:2},{M}_t^{1:2},Z_t^{er},{Y}_{t}),\\
\label{commonevol2}{Z}_{t+1} &=\zeta_{t+1}({P}_{t^+}^{1:2},{U}_t^{1:2},X_{t+1}^{0:2})
\end{align}
where $\zeta_{t+1}$ and $\zeta_{t^+}$ are fixed transformations.
\item \emph{Private Information Evolution: } The private information evolves as
\begin{align}
\label{privevol}{P}^{i}_{t^+} &= \xi_{t^+}^{i}({P}_t^{1:2},{M}_t^{1:2},Z_t^{er},{Y}_{t})\\
\label{privevol2}{P}^{i}_{t+1} &= \xi_{t+1}^{i}({P}_{t^+}^{1:2},{U}_t^{1:2},{X}_{t+1}^{0:2})
\end{align}
where $\xi_{t^+}^{i}$ and $\xi_{t+1}^{i}$ are fixed transformations {and $i=1,2$}.
\end{enumerate}
\end{assumption}
Due to the nestedness condition in Assumption \ref{infoassump}, the team is always more-informed than the adversary. Scenarios where the adversary has some private information are beyond the scope of this paper. The third and fourth conditions in Assumption \ref{infoassump} on the evolution of common and private information are very mild \cite{nayyar2013decentralized,nayyar2014common} and most information structures of interest satisfy these conditions.

\textbf{Strategy optimization problem:} At time $t$, the system incurs a cost $c_t(X^0_t,X_t,U_t,U^a_t)$ that depends on the global state, the team's state, control actions of both agents and the adversary's action. Whenever agents decide to share their states with each other, they incur a state-dependent  cost $\rho(X_t^0,X_t)$. 
The system runs for a time horizon $T$. The total expected cost over the time horizon $T$ associated with a strategy profile $((f,g), g^a)$ is:
\begin{align}
&J((f,g), g^a)=\label{eq:cost2}\\
&\ee^{((f,g), g^a)}\Big[\sum_{t=1}^{T} c_t(X_t^0,{X}_t,{U}_t,U_t^a)+\rho(X_t^0,X_t)\mathds{1}_{\{M^{or}_t=1\}}\Big].\notag
\end{align}
The objective of the team is to find communication and control strategies $(f,g)$ for the team in order to minimize the worst-case expected total expected cost $\max_{g^a} J((f,g), g^a)$.
This min-max optimization problem can be viewed as a zero-sum game between the team and the adversary. We denote this zero-sum game with Game $\gm{G}$. We denote the min-max value of this game $\gm{G}$ with $S^u(\gm{G})$, i.e.,
\begin{align}
    S^u(\gm{G}) = \min_{(f,g)}\max_{g^a}J((f,g), g^a).\label{costdef}
\end{align}
\begin{remark}
The strategy spaces of all the players (agents in the team and the adversary) are compact and the cost $J(\cdot)$ is continuous in $f,g,g^a$. Hence, we can conclude using Berge's maximum theorem \cite{guide2006infinite} that there exist strategies that achieve the maximum and minimum in \eqref{costdef}.
\end{remark}


\subsection{Examples of Information Structures Satisfying Assumption \ref{infoassump}}\label{infoexamples}
\subsubsection{Maximum information}\label{fullinfo} Consider the case where the adversary's information at time $t$ and $t^+$ is given by
    \begin{align}
         I_{t}^a & = \cteam_t \nonumber\\
        I_{t^+}^a & = \cteam_{t^+}.\nonumber
    \end{align}
In this case, the adversary has access to all the information it can while satisfying Assumption \ref{infoassump}. Thus, the common information $C_t = \cteam_t$ (resp. for $t^+$). The private information for agent $i$ at time $t$ is given by $P_t^i = \{X_{1:t}^i,U_{1:t-1}^i\}$. This information structure models scenarios in which agents in the team do not use any form of encryption and any communication that happens between them can be observed by the adversary.

\subsubsection{Encrypted Communication with Global State Information}\label{fullencrypt}
Consider the following information structure for the adversary
\begin{align}
    I^a_{t} &= \{X_{1:t}^0,U_{1:t-1}^a,M^{1,2}_{1:t-1}\}\\
    I_{t^+}^a &= \{X_{1:t}^0,U_{1:t-1}^a,M^{1,2}_{1:t}\}\\
    Y_t&=0.
\end{align}
This information structure models scenarios in which agents in the team have the capability to encrypt their messages. Since the adversary's observation $Y_t$ is a constant, it has no knowledge about the messages exchanged by the team. The adversary however knows whether or not communication was initiated by the agents. 
The private information for agent $i$ at time $t$ is given by $P_t^i = \{X_{1:t}^i,U_{1:t-1}^i,Z_{1:t-1}^{er}\}$.

\subsubsection{Imperfect Encryption with Global State Information}\label{imperfectencrypt}
\begin{align}
    I^a_{t}  &=\{X_{1:t}^0,U_{1:t-1}^a,M^{1,2}_{1:t-1},Y_{1:t-1}\}\\
    I^a_{t^+}  &=\{X_{1:t}^0,U_{1:t-1}^a,M^{1,2}_{1:t},Y_{1:t}\}
\end{align}
This information structure is very similar to the one discussed above except that the encryption mechanism used by the agents may be imperfect. 




\section{Preliminary Results and Simplified Game $\gm{G}_s$}\label{Preliminary_Results}
In this section we show that agents in the team can ignore parts of their information without losing optimality. This removal of information narrows the search for optimal strategies to a class of simpler strategies and is a key step in our approach for finding optimal strategies.


Let us define the team's \emph{common private information} $D_t$ before communication at time $t$ and $D_{t^+}$ after communication at time $t^+$ as
\begin{align}
    D_t &:= P_t^1 \cap P_t^2\\
    D_{t^+} &:= P_{t^+}^1 \cap P_{t^+}^2.
\end{align}
The variables $C_t,D_t$ (resp. $C_{t^+},D_{t^+}$) constitute the \emph{team's common information} at time $t$ (resp. $t^+$), i.e.,
\begin{align}
    C_t \cup D_t &= I_t^1 \cap I_t^2 = \cteam_t\\
    C_{t^+} \cup D_{t^+} &= I_{t^+}^1 \cap I_{t^+}^2 = \cteam_{t^+}.
\end{align}
Notice that $C_t$ and $D_t$ depend on the adversary's information structure. However, since the team's information structure is fixed, $C_t,D_t$ combined do not depend on the adversary's information structure.
The following lemma establishes a key conditional independence property that will be critical for our analysis. 
\begin{lemma}[Conditional independence property] \label{LEM:INDEPEN2}
At any time $t$, the two agents'  local states  and control actions   are conditionally independent given the team's common information $(C_t,D_t)$ (before communication) or $C_{t^+},D_{t^+}$ (after communication). That is, if $c_t,d_t, c_{t^+},d_{t^+}$ are the realizations of the common information and common private information before and after communication respectively, then for any realization ${x}_{1:t}, {u}_{1:t-1}$ of states and actions, we have
\begin{equation}\label{eq:indepen}
    \prob({x}_{1:t},{u}_{1:t-1}|c_t,d_t)=\displaystyle\prod_{i=1}^{2} \prob(x^{i}_{1:t},{u}^{i}_{1:t-1}|c_t,d_t), 
    \end{equation}
    \begin{equation}\label{eq:indepen2}
      \prob({x}_{1:t},{u}_{1:t}|c_{t^+},d_{t^+})=\displaystyle\prod_{i=1}^{2} \prob(x^{i}_{1:t},{u}^{i}_{1:t}|c_{t^+},d_{t^+}).  
    \end{equation}
    Further, $\prob(x^{i}_{1:t},{u}^{i}_{1:t-1}|c_t,d_t)$ and $ \prob({x}^i_{1:t},{u}^i_{1:t}|c_{t^+},d_{t^+})$ depends on only on agent $i$' strategy.
\end{lemma}    
\begin{proof}
The proof of this lemma is very similar to the proof of Lemma 1 in \cite{sudhakara2021optimal}. For a detailed proof, see Appendix \ref{proof:CI1} in \cite{extended}.
\end{proof}

The following proposition shows that agent $i$ at time $t$ and $t^+$ can ignore its past states and actions, i.e. $X^i_{1:t-1}$ and $U^i_{1:t-1}$, without losing optimality. This allows agents in the team to use simpler strategies where the communication and control decisions are functions only of the current state and the team's common information.
\begin{proposition} \label{PROP:ONE2}
 Agent $i$, $i=1,2,$ can restrict itself to strategies of the form below
\begin{equation} \label{eq:structure1}
   M^i_t \sim \bar{f}^i_t(X^i_{t},C_t,D_t)
\end{equation}
\begin{equation}  \label{eq:structure2}
   U^i_t \sim \bar{g}^i_t(X^i_{t},C_{t^+},D_{t^+})
\end{equation}
without loss of optimality.  In other words, at time $t$ and $t^+$, agent $i$ does not need the past local states and actions, $X^i_{1:t-1},U^i_{t-1}$, for making optimal decisions.
\end{proposition}
\begin{proof}
See Appendix \ref{proponeproof}.
\end{proof}

Proposition \ref{PROP:ONE2} leads to a simplified game in which the information used by the players in the team is substantially reduced. We will refer to this game as Game $\gm{G}_s$. Game $\gm{G}_s$ has the same dynamics and cost model as Game $\gm{G}$. The key difference between these two games lies in the team's information structure and strategy spaces. In Game $\gm{G}_s$, the information used by player $i$ in the team at time $t$ and $t^+$ respectively is
\begin{align}
    I_t^i &= \{X_t^i\} \cup D_t \cup C_t \label{redinft}\\
    I_{t^+}^i &= \{X_t^i\} \cup D_{t^+}\cup C_{t^+}.\label{redinftp}
\end{align}
Therefore, the common information in the simplified game $\gm{G}_s$ is the same as in the original game $\gm{G}$. In the simplified game $\gm{G}_s$, the private information\footnote{With a slight abuse of notation, we use the same letter for denoting private information in both games $\gm{G}$ and $\gm{G}_s$.} $P_t^i = X_t^i \cup D_t$.

\begin{corollary}
If $(f^*,g^*)$ is a min-max strategy in Game $\gm{G}_s$, then it is a min-max strategy in Game $\gm{G}$. Further, the min-max values of games $\gm{G}$ and $\gm{G}_s$ are identical.
\end{corollary}
Henceforth, we make the following mild assumption on the information structure of agents in the simplified game $\gm{G}_s$.  It can be easily verified that the simplified games corresponding to all the models discussed in Section \ref{infoexamples} satisfy this assumption.
\begin{assumption}\label{infoassum2}
The information structure in the simplified game $\gm{G}_s$ with reduced private information satisfies Assumption \ref{infoassump}.
\end{assumption}
\begin{remark}
The reduced information in equations \eqref{redinft} and \eqref{redinftp} is \emph{unilaterally sufficient information} (see Definition 2.4 in \cite{tang2021games}) for each player in the team. Proposition \ref{PROP:ONE2} can alternatively be shown using the concept of unilaterally sufficient information and Theorem 2.6 in \cite{tang2021games}. 
\end{remark}


\section{Dynamic Program Characterization of a Min-max Strategy}\label{CoordinatorResults}
It was shown in \cite{kartikarxiv} that for certain zero-sum game models with a special structure, a virtual game $\gm{G}_e$ can be constructed based on the simplified Game $\gm{G}_s$, and this virtual game can be used to obtain the min-max value and a min-max strategy for the minimizing team. In our game model described in Section \ref{sec:problem_formulation}, the adversary does not have any private information at any given time and hence, this model can be viewed as a special case of the game model described in paragraph ($a$), Section IV-A of \cite{kartikarxiv}. Therefore, we can use the result in \cite{kartikarxiv} to obtain the min-max value and a min-max strategy for our original Game $\gm{G}$. The virtual game $\gm{G}_e$ involves the same underlying system model as in game $\gm{G}_s$. The main differences among games $\gm{G}_s$ and $\gm{G}_e$ lie in the manner in which the actions used to control the system are chosen. In the virtual game $\gm{G}_e$, 
all the players in the team of game $\gm{G}_s$ are replaced by a virtual player (referred to as virtual player $b$) and the adversary is replaced by a virtual player (referred to as virtual player $a$). These virtual players in Game $\gm{G}_e$ operate as described in the following sub-section.

\subsection{Virtual Game $\gm{G}_e$}\label{expgame}

Consider virtual player $a$ associated with the adversary. At each time $t^+$, virtual player $a$ selects a distribution $\Gamma^a_t$ over the space $\mathcal{U}_t^{a}$. The set of all such mappings is denoted by $\mathcal{B}_t^{a} \doteq \Delta \U_t^{a}$. Consider virtual player $b$ associated with the team. At each time $t$ and for each $i = 1,2$, virtual player $b$ selects a function $\Gamma^{i}_t$ that maps private information $P^{i}_t$ to a distribution $\delta {M}_t^{i}$ over the space $\{0,1\}$. Thus, $\delta M_t^{i} = \Gamma_t^{i}(P_t^{i})$. The set of all such mappings is denoted by $\mathcal{B}_t^{i} \doteq \mathcal{F}(\P_t^{i},\Delta\{0,1\})$. We refer to the tuple $ \Gamma_t \doteq (\Gamma_t^{1},\Gamma_t^{2})$ as virtual player $b$'s \emph{prescription} at time $t$. {The set of all possible prescriptions for virtual player $b$ at time $t$ is denoted by $\mathcal{B}_t \doteq \mathcal{B}_t^{1}\times\mathcal{B}_t^{2}$.} At each time $t^+$ and for each $i = 1,2$, virtual player $b$ selects a function $\Lambda^{i}_t$ that maps private information $P^{i}_{t^+}$ to a distribution $\delta {U}_{t^+}^{i}$ over the space $\mathcal{U}_t^i$. Thus, $\delta U_t^{i} = \Lambda_t^{i}(P_{t+}^{i})$. The set of all such mappings is denoted by $\mathcal{B}_{t^+}^{i} \doteq \mathcal{F}(\P_{t^+}^{i},\Delta\mathcal{U}_t^i)$. We refer to the tuple $ \Lambda_t \doteq (\Lambda_t^{1},\Lambda_t^{2})$ as virtual player $b$'s \emph{prescription} at time $t^+$. {The set of all possible prescriptions for virtual player $b$ at time $t^+$ is denoted by $\mathcal{B}_{t^+} \doteq \mathcal{B}_{t^+}^{1}\times\mathcal{B}_{t^+}^{2}$.}
Once virtual players select their prescriptions at times $t$ and $t^+$, the corresponding actions are generated as
\begin{align}
M^i_t& = \rand(\{0,1\}, \Gamma^i_t(P_t^i),K_t^i)\\
U^i_t& = \rand(\mathcal{U}_t^i,\Lambda^i_t(P_{t^+}^i),K_{t^+}^i)\\
U_t^a &= \rand(\mathcal{U}_t^a,\Gamma_t^a,K_{t^+}^a).
\end{align}



In virtual game $\gm{G}_e$, virtual players' information $I_t^v$ at time $t$ comprises of the common information ${C}_t$ and the past prescriptions of both players $\Gamma_{1:t-1},\Gamma_{1:t-1}^a,\Lambda_{1:t-1}$. At time $t$, Virtual player $b$ selects its prescription according to a control law $\chi_t^b$, i.e. $\Gamma_t = \chi_t^b(I_t^v)$. Note that at time $t$, Virtual player $a$ does not take any action. At time $t^+$, the virtual players information $I_{t^+}^v$ comprises of $C_{t^+}$ and all the past prescriptions of both players $\Gamma_{1:t},\Gamma_{1:t-1}^a,\Lambda_{1:t-1}$. Virtual player $a$ selects its prescription according to a control law $\chi_t^a$, i.e., $\Gamma_t^a = \chi_t^a(I_{t^+}^v)$ and virtual player $b$ selects its prescription according to a control law $\chi_{t^+}^b$, i.e. $\Lambda_t = \chi_t^b(I_{t^+
}^v)$.
 For virtual player $a$, the collection of control laws over the entire time horizon ${\chi}^a = (\chi_1^a,\dots,\chi_T^a)$ is referred to as its control strategy. Similarly for virtual player $b$. Let   $\mathcal{H}_t^a$ be the set of all possible control laws for virtual player $a$ at time $t$  and let  $\mathcal{H}^a$ be the set of all possible control strategies for virtual player $a$, i.e. $\mathcal{H}^a = \mathcal{H}_1^a \times \dots \times \mathcal{H}_T^a$. For virtual player $b$, the collection of control laws over the entire time horizon ${\chi}^b = (\chi_1^b,\chi_{1^+}^b\dots,\chi_T^b,\chi_{T^+}^b)$ is referred to as its control strategy. Let   $\mathcal{H}_t^b$ (resp. $\mathcal{H}_{t^+}^b$)  be the set of all possible control laws for virtual player $b$ at time $t$ (resp. $t^+$) and let  $\mathcal{H}^b$ be the set of all possible control strategies for virtual player $b$. 
The total cost associated with the game for a strategy profile $({\chi}^a,{\chi}^b)$ is
\begin{align}\label{eq:virtualcost2}
&\mathcal{J}(\chi^a, \chi^b)=\\
&\ee^{(\chi^a,\chi^b)}\Big[\sum_{t=1}^{T}c_t(X_t^0,{X}_t,{U}_t,U_t^a)+\rho(X_t^0,X_t)\mathds{1}_{\{M^{or}_t=1\}}\Big].\notag
\end{align}
where the functions $c_t$ and $\rho$ are the same as in games $\gm{G}$ and $\gm{G}_s$.
In this virtual game, virtual player $a$ aims to maximize the cost while virtual player $b$ aims to minimize the cost. The upper value of Game $\gm{G}_e$ is denoted by $S^u(\gm{G}_e)$.

\subsection{Common Information Belief and the Dynamic Program}

\subsubsection{Common Information Belief}  Before communication at time $t$, the CIB is given as:
\begin{align}
    & {\Pi_t(x^0,x,d)=\prob[X^0_t=x^0,X_t=x, D_t =d|I_{t}^{v}}].
\end{align}
After the communication decisions are made and $Z^{er}_t$ is realized, the CIB is given as:
\begin{align}
    &\Pi_{t^+}(x^0,x,d)=\prob[X^0_t=x^0,X_t=x, D_{t^+}=d|I_{t^+}^v].
\end{align}
The CIB satisfies two key properties: (i) the CIB can be computed without using the virtual players' strategies $\chi^a$ and $\chi^b$; (ii) since the adversary does not have any private information at any given time, the CIB does not depend on the adversary's prescriptions (see Section IV-A and Appendix VI of \cite{kartikarxiv}). This can be stated formally as the following lemma.
\begin{lemma}\label{LEM:UPDATE}
$\Pi_1(x_1^0,x_1,d_1)$ is the belief $\prob(X_1^0 = x_1^0,{X_1} =x_1,D_1 = d_1)$  and for each $t \geq 1$, 
\begin{equation}
    \Pi_{t^+}=\eta_t( \Pi_{t},\Gamma_t,Z_{t^+}),
\end{equation}
\begin{equation}
    \Pi_{t+1}=\beta_t( \Pi_{t^+},\Lambda_{t},Z_{t+1}),
\end{equation}
where   $\eta_t, \beta_t$ are fixed transformations derived from the system model using Bayes' rule (see Appendix VI of \cite{kartikarxiv}). 
\end{lemma}

We now describe the dynamic program that provides us with the value of the game $\gm{G}$ and an algorithm to compute a min-max strategy for the team.
\subsubsection{Dynamic Program}\label{dp}
Define the value function $V_{T+1}(\pi):=0$ for all $\pi$ at time $T+1$. The cost-to-go functions $w_t$  (resp. $w_{t^+}$) and value functions $V_t$ (resp. $V_{t^+}$) for  $t=T, \ldots, 2, 1,$  are defined as follows:
 \begin{align}
 &w_{t^+}(\pi,\lambda,\gamma^a) := \notag \\
 & \E\Big[{c}_t(X_t^0,X_t,U_t,U_t^a) +V_{t+1}(\beta_t( \pi,\lambda,Z_{t+1}))\mid \pi,\lambda,\gamma\Big],\nonumber\\
  &V_{t^+}(\pi) := \min_{\lambda}\max_{\gamma^a} w_{t^+}(\pi),\label{vtp}\\
    &w_{t}(\pi,\gamma) := \notag\\
 &\E\Big[\rho(X_t^0,X_t)\mathds{1}_{\{M^{or}_t=1\}}+ V_{t^+}(\eta_t( \pi^{1,2},\gamma, Z_{t^+}))\mid \pi,\gamma\Big],\notag\\
&V_{t}(\pi) :=\min_{\gamma} w_{t}(\pi,\gamma).\label{vt}
\end{align}
Let $\Xi_t(\pi)$ (resp. $\Xi_{t^+}(\pi)$) be a minimizer (resp. minmaximizer) of the cost-to-go function in \eqref{vt} (resp. \eqref{vtp}).

{\begin{theorem}\label{onesidevalue}
The min-max value of games $\gm{G}$, $\gm{G}_s$ and $\gm{G}_e$ are identical, i.e., we have $S^u(\gm{G}) = S^u(\gm{G}_e) =  \E[V_1(\Pi_1)].$ Further, the strategy pair $f^*,{g}^{*}$ described by Algorithm \ref{alg:example} is a min-max strategy for the team in the original game $\gm{G}$.
\end{theorem}}
\begin{proof}
Because of our assumption on the information structure of Game $\gm{G}_s$ (Assumption \ref{infoassum2}), the evolution of CIB in Game $\gm{G}_e$ does not depend virtual player $a$'s prescription. This property allows us to use Theorems 4 and 5 in \cite{kartikarxiv} and obtain our result.
\end{proof}
The dynamic program is helpful for characterizing the min-max value and a min-max strategy in a general setting. However, solving the dynamic program involves computational challenges. The main cause of these challenges is that the private information ($X_t^i \cup D_t$) space can be very large even after the private information reduction in the simplified game $\gm{G}_s$. For instance, $D_t = Z_{1:t-1}^{er}$ in the model described in Section \ref{fullencrypt}. In the following sub-sections, we discuss some special cases in which the private information is small or can be reduced further to a manageable size. Once the private information has been reduced sufficiently, one can use the computational methodology discussed in Appendix X of \cite{kartikarxiv} to solve the dynamic program.

\begin{algorithm}[tb]
  \caption{Strategies $f^{i*},g^{i*}$ for Player $i$ in the Team}
  \label{alg:example}
\begin{algorithmic}
    \STATE Input: $\Xi_t(\pi), \Xi_{t^+}(\pi)$ obtained from DP for all $t$ and all $\pi$
  \FOR{$t=1$ {\bfseries to} $T$}
  \STATE \textbf{Before communication:}
  \STATE Current information: $C_t,P_t^{i}$
  \COMMENT{where $C_{t} = \{C_{(t-1)^+},Z_t\}$}
  \STATE Update CIB $\Pi_{t} = \beta_{t-1}(\Pi_{(t-1)^+}, \Xi_{(t-1)^+}^1(\Pi_{t-1^+}),Z_{t})$ \COMMENT{If $t=1$, Initialize CIB $\Pi_t$ using $C_1$}
  \STATE Get prescription $\Gamma_t = (\Gamma_t^{1},\Gamma_t^{2}) = \Xi_t(\Pi_t)$ 
  \STATE Get distribution $\delta M_t^{i} = \Gamma_t^{i}(P_t^{i}) $ and select action $M_t^{i} = \rand(\{0,1\},\delta M_t^{i},{K}_t^{i})$
  \STATE \textbf{After communication decisions are made:}
  \STATE Current information: $C_{t^+},P_{t^+}^{i}$
  \COMMENT{where $C_{t^+} = \{C_{t},Z_{t^+}\}$}
  \STATE Update CIB $\Pi_{t^+} = \eta_t(\Pi_{t}, \Xi_{t}^1(\Pi_t),Z_{t^+})$
  \STATE Get prescription $\Lambda_t = (\Lambda_t^{1},\Lambda_t^{2}) = \Xi_{t^+}(\Pi_{t^+})$ 
  \STATE Get distribution $\delta U_t^{i} = \Lambda_t^{i}(P_t^{i}) $ and select action $U_t^{i} = \rand(\mathcal{U}_t^{i},\delta U_t^{i},{K}_{t^+}^{i})$

  \ENDFOR
\end{algorithmic}
\end{algorithm}

\subsection{Communication without Encryption}\label{commencryptsec}
Consider the information structure in Section \ref{fullinfo} in which the agents do not encrypt their information. In this case, $D_t$ is empty and therefore, the only private information agent $i$ uses at time $t$ and $t^+$ is $X_t^i$. Due to this reduced private information space, the prescription space in the virtual game $\gm{G}_e$ is substantially smaller. Further, the CIB at time $t$ is formed only on the current state $X_t$ and thus, can be updated easily. The smaller prescription space, belief space and simpler belief update significantly improve the computational tractability of the dynamic program in  Section \ref{dp}. The CIB update rules for this information structure will be used in proving other results and hence, we denote these update rules with $\bar{\eta}_t$ and $\bar{\beta}_t$.

\subsection{Communication with Encryption}
In this section, we consider the information structure described in Section \ref{imperfectencrypt}. In this information structure, agents in the team can encrypt their information perfectly or imperfectly. At any given time, the adversary can observe whether or not communication was initiated by each agent in the team and subsequently, the imperfectly encrypted message that was exchanged between the agents. Further, we assume that the adversary can observe whether or not the agents had a successful communication over the erasure channel. Let $E_t$ be a Bernoulli variable such that $E_t = 1$ if and only if successful communication occurred at time $t$. Note that $E_t = 1-\mathds{1}_{\varnothing}(Z_t^{er})$ and therefore, can be viewed as a part of the adversary's observation $Y_t$ (see \eqref{advobs}).

Let us define the time of last communication at time $t$ as
\begin{align}
    L_t &= \max\{\tau : \tau < t, E_\tau = 1\}\\
    L_{t^+} &= \max\{\tau : \tau \leq t, E_\tau = 1\}.
\end{align}
Here, the maximum of an empty set is considered to be $-\infty$. Note that $L_{t+1} = L_{t^+}$ and that the adversary can compute $L_t$ and $L_{t^+}$ using its information.

\begin{proposition}\label{privateredencrypt}
There exists a min-max strategy of the form
\begin{align}
     M^i_t &\sim  f_t^i(X_t^i,X_{L_t},C_t)\\
      U^i_t &\sim {g}^i_t(X^i_{t},X_{L_{t^+}},C_{t^+})
\end{align}
\end{proposition}
\begin{proof}
See Appendix \ref{beliefpropproof}.
\end{proof}
The main consequence of Proposition \ref{privateredencrypt} is that agent $i$'s private information at time $t$ (resp. $t^+$) is reduced from $X_t^i,Z_{1:t-1}^{er}$ (resp. $X_t^i,Z_{1:t}^{er}$) to $X_t^i,X_{L_t}$ (resp. $X_t^i,X_{L_{t^+}}$). As discussed in the previous sub-section, this reduction leads to the simplification of the dynamic program described in Section \ref{dp}.

\section{Conclusions}
We considered a zero-sum game between a team of two agents and a malicious agent. The agents can strategically decide at each time whether to share their private information with each other or not. The agents incur a cost whenever they communicate with each other and the adversary may eavesdrop on their communication. Under certain assumptions on the system dynamics and the information structure of the adversary, we characterized a min-max control and communication strategy for the team using a common information belief based min-max dynamic program. For certain specialized information structures, we proved that the agents in the team can ignore a large part of their private information without losing optimality. This reduction in private information substantially simplifies the dynamic program and hence, improves computational tractability.

 \bibliographystyle{IEEEtran}
\bibliography{refs}


\appendices
\section{Proof of Lemma \ref{LEM:INDEPEN2}}
\label{proof:CI1}
We prove the lemma by induction. 

\textbf{Induction hypothesis at time $t$:} Equation \eqref{eq:indepen} holds at time $t$. 

At $t=1$, before communication decisions are made, our induction hypothesis is trivially true since the team's common information at this point is the global state $X_1^0$ and the agents' initial states $X_1^1,X_1^2$ are independent given the global state $X_1^0 = x_1^0$ for any $x_1^0 \in \mathcal{X}^0$.


\textbf{Induction step:} 
Using the induction hypothesis at time $t$, we first prove \eqref{eq:indepen2} at time $t^+$. In order to do so, it suffices to show that the left hand side of \eqref{eq:indepen2} can be factorized as follows:
\begin{align}
&\prob({x}_{1:t},{u}_{1:t}|c_{t^+},d_{t^+})\notag\\
&=\zeta^1({x}^1_{1:t},{u}^1_{1:t},c_{t^+},d_{t^+})\zeta^2({x}^2_{1:t},{u}^2_{1:t},c_{t^+},d_{t^+}),\label{factor}
\end{align}
 where $\zeta^1$ and $\zeta^2$ are some real-valued mappings with $\zeta^i$ depending only on agent $i$'s strategy. We now factorize the joint distribution below. Recall that {$c_{{t}^+},d_{{t}^+} = (c_{t},d_{t},z^{er}_{t},m_{t},y_t)$}, $I^i_{{t}^+}=(x^0_{1:t},u^a_{1:t-1},x^i_{1:t},u^i_{1:t-1},z^{er}_{1:t},m_{1:t},y_{1:t})$ and $I^i_{{t}}=(x^0_{1:t},u^a_{1:t-1},x^i_{1:t},u^i_{1:t-1},z^{er}_{1:t-1},m_{1:t-1},y_{1:t-1})$. The left hand side of \eqref{eq:indepen2} for $t^+$ can be written as
\begin{align}
&\frac{\prob(x_{1:t},u_{1:t},z^{er}_{t},m_{t},y_t| c_{t}, d_{t})}{\prob(z^{er}_{t},m_{t},y_t | c_{t},d_{t})}=\notag\\
 &\frac{\prob(u_{t}|x_{1:t},u_{1:t-1},c_{t^+},d_{t^+})\prob(y_t,z^{er}_{t}|x_{1:t},u_{1:t-1},c_{t},d_{t},m_t)}{\prob(z^{er}_{t},m_{t},y_t | c_{t},d_{t})}\notag \\
 &\times \prob(m_{t}|x_{1:t},u_{1:t-1},c_{t},d_{t})\prob(x_{1:t},u_{1:t-1}|c_{t},d_{t})\notag\\
 &=\left({\prob(m^1_{t}|I^1_{{t}})}{\prob(u^1_{t}|I^1_{{t}^+})}\prob(x^1_{1:t},u^1_{1:t-1}|c_{t},d_{t})\right)\notag\\
 &\times\left({\prob(m^2_{t}|I^2_{{t}})}{\prob(u^2_{t}|I^2_{{t}^+})}\prob(x^2_{1:t},u^2_{1:t-1}|c_{t},d_{t})\right) \notag\\
&\quad\times\frac{\prob(y_t,z^{er}_{t}|x_{1:t},u_{1:t-1},c_{t},d_{t},m_{t})}{\prob(y_t,z^{er}_{t},m_{t} | c_{t},d_{t})},\label{rhss2:inductive}
\end{align}
where the last equality follows from induction hypothesis at time $t$. Further, we have
\begin{align}
\label{3cases}&\prob(y_t,z^{er}_{t}|x_{1:t},u_{1:t-1},c_{t},d_{t},m_{t})\\
&= \prob(y_t\mid z_t^{er},m_t,x_t^0)\times\notag\\
\notag&
\begin{cases}
1 & \text{if $m_{t} = (0,0)$, $z^{er}_{t}=\phi$} \\
{p_e(x_t^0)} & \text{if $m_{t} \neq (0,0)$, $z^{er}_{t}=\phi$}\\
(1-{p_e(x_t^0)})\mathds{1}_{\{x_{t}=(\tilde{x}^1_{t}, \tilde{x}^2_{t})\}} & \text{if $m_{t} \neq (0,0)$, $z^{er}_{t}=(\tilde{x}^1_{t},\tilde{x}^2_{t})$},
\end{cases}
\end{align}
which can clearly be factorized. From equations \eqref{rhss2:inductive} and \eqref{3cases}, the joint distribution $\prob({x}_{1:t},{u}_{1:t}|c_{t^+},d_{t^+})$ can be factorized as in \eqref{factor} and thus \eqref{eq:indepen2} holds at time $t^+$.
Using this result, we now show that our induction hypothesis holds at time $t+1$. Recall that {$c_{t+1},d_{t+1} = (c_{t^+},d_{t^+},x_{t+1}^0,u_{t}^a)$}. At time $t+1$, before 
communication decisions are made, the left hand side of equation \eqref{eq:indepen} can be written as
{\begin{align}
 &\prob(x_{1:t+1},u_{1:t}|c_{t+1},d_{t+1})
 = \frac{\prob(x_{1:t+1},u_{1:t},x_{t+1}^0,u_t^a|c_{t^+},d_{t^+})}{\prob(x_{t+1}^0,u_t^a|c_{t^+},d_{t^+})}\\
 &=\frac{\prob(x_{t+1}|x_{1:t},u_{1:t},c_{t+1},d_{t+1})\prob(x_{t+1}^0|x_{1:t},u_{1:t},u_{t}^a,c_{t^+},d_{t^+})}{\prob(x_{t+1}^0,u_t^a|c_{t^+},d_{t^+})}  \notag\\
 &\times\prob(u_{t}^a|x_{1:t},u_{1:t},c_{t^+},d_{t^+})\prob(x_{1:t},u_{1:t}|c_{t^+},d_{t^+})\notag\\
 &=\frac{\prob(x_{t+1}|x_{1:t},u_{1:t},c_{t+1},d_{t+1})}{\prob(x_{t+1}^0,u_t^a|c_{t^+},d_{t^+})}\times\notag\\
 &\prob(x_{t+1}^0|u_{t}^a,c_{t^+},d_{t^+}) \prob(u_{t}^a|c_{t^+},d_{t^+})\prob(x_{1:t},u_{1:t}|c_{t^+},d_{t^+})\label{ind1}  \\
&=\prob(x^{2}_{t+1}|x^{1}_{t+1},x_{1:t},u_{1:t},c_{t+1},d_{t+1})\times\notag\\
&\prob(x^1_{t+1}|x_{1:t},u_{1:t},c_{t+1},d_{t+1})\prob(x_{1:t},u_{1:t}|c_{t^+},d_{t^+}) \notag \\
&=\prob(x^{2}_{t+1}|x_t^0,x_t^2,u_t^2)\times\notag\\
&\prob(x^1_{t+1}|x_t^0,x_t^1,u_t^1)\prob(x_{1:t},u_{1:t}|c_{t^+},d_{t^+}) \label{ind2} \\
 &=\left[\prob(x^{1}_{t+1}|x_t^0,x^1_{t},u^1_{t})\prob(x^1_{1:t},u^1_{1:t}|c_{t^+},d_{t^+})\right]\times\notag\\
 &\left[\prob(x^{2}_{t+1}|x_t^0,x^2_{t},u^2_{t})\prob(x^2_{1:t},u^2_{1:t}|c_{t^+},d_{t^+})\right]. \label{lnn1:inductive}
\end{align}}
{Here, \eqref{ind1} and \eqref{ind2} are consequences of the system dynamics in \eqref{dyna}. The last equation \eqref{lnn1:inductive} follows from the equation \eqref{eq:indepen2} at time $t^+$.} Using the factored form of $\prob(x_{1:t+1},u_{1:t}|c_{t+1},d_{t+1})$ in \eqref{lnn1:inductive}, we can conclude that our induction hypothesis holds at time $t+1$.

Therefore, by induction, we can conclude that equations \eqref{eq:indepen} and \eqref{eq:indepen2} hold at all times.
 \section{Proof of Proposition \ref{PROP:ONE2}}\label{proponeproof}
We will prove the lemma using the following claim.
\begin{claim}\label{claim:one}
Consider any arbitrary strategy pair $f,g$ for the team. Then there exists a strategy pair $\bar{f},\bar{g}$ for the team such that, for each $t$, $\bar{f}^{i}_t$ (resp. $\bar{g}^{i}_t$) are functions of $X_t^i$ and $C_t,D_t$ (resp. $C_{t^+},D_{t^+}$) and 
\[
J((\bar{f},\bar{g}),g^a) = J((f,g), g^a), ~~\forall g^a \in \mathcal{G}^a.
\]

\end{claim}

Suppose that the above claim is true. Let $(f',g')$ be a min-max strategy for the team.
Due to Claim \ref{claim:one}, there exists a strategy $\bar{f'},\bar{g'}$ for the team such that, for each $t$, $\bar{f'}^{i}_t$ (resp. $\bar{g'}^{i}_t$) are functions of $X_t^i$ and $C_t,D_t$ (resp. $C_{t^+},D_{t^+}$) and 
\[
J((\bar{f'},\bar{g'}),g^a) = J((f',g'), g^a),
\]
\emph{for every strategy $g^a \in \mathcal{G}^a$.} Therefore, we have that 
\begin{align*}
\sup_{g^a \in \mathcal{G}^a}J((\bar{f'},\bar{g'}),g^a) &= \sup_{g^a \in \mathcal{G}^a}J((f',g'),g^a) \\
&= \inf_{f,g}\sup_{g^a \in \mathcal{G}^a}J((f,g),g^a).
\end{align*} 
Thus, $(\bar{f}',\bar{g}')$ is a min-max strategy for Team 1 wherein Player 1 uses only the current state and Player 2's information.

It suffices to prove Claim \ref{claim:one} for the information structure in Section \ref{fullinfo}. In this information structure, the adversary has the maximum possible information. Given a strategy $\hat{g}^{a}$ under any other information structure, there is an equivalent strategy $g^a$ under the maximal information structure in Section \ref{fullinfo}. Therefore, the set of all adversary's strategies in the maximal information structure covers the adversary's strategy space under any other information structure.

The proof of Claim \ref{claim:one} is similar to the proof of Lemma 3 in Appendix IX of \cite{kartikarxiv}. However, there is an important difference in the construction of $(\bar{f}',\bar{g}')$. For this construction, the conditional independence in Lemma \ref{LEM:INDEPEN2} is crucial. 

\noindent{\emph{Proof of Claim \ref{claim:one}:}} We now proceed to prove Claim \ref{claim:one} for the information structure in Section \ref{fullinfo}.

Consider any arbitrary strategy $f,g$ for the team. Let $\iota_t^a = \{x_{1:t}^0,u_{1:t-1}^{a},z_{1:t-1}^{er},m_{1:t-1},y_{1:t-1}\}$ be a realization of the adversary's information $I_t^a$ at time $t$.
Define the distribution $\Psi_t(\iota_t^a)$ over the space $(\prod_{\tau = 1}^t\prod_{i=1,2}\X^i_{\tau} \times\prod_{\tau = 1}^{t-1}\prod_{i=1,2}\U_{\tau}^{i})\times \{0,1\}^2$ as follows:
\begin{align*}
&\Psi_t(\iota_t^a; x_{1:t},u_{1:t-1},m_t) \\
&\doteq \Py^{f,g,h^a}[X_{1:t},U_{1:t-1},M_t = (x_{1:t},u_{1:t-1},m_t) \mid \iota_t^a],
\end{align*}
if $\iota_t^a$ is \emph{feasible}, that is $\Py^{f,g,h^a}[I_t^a = \iota_t^a] > 0$, under the \emph{open-loop} strategy $h^a \doteq (u^a_{1:t-1})$ for the adversary. Otherwise, define $\Psi_t(\iota_t^a; x_{1:t},u_{1:t-1},m_t)$ to be the uniform
 distribution over the space $(\prod_{\tau = 1}^t\prod_{i=1,2}\X^i_{\tau} \times\prod_{\tau = 1}^{t-1}\prod_{i=1,2}\U_{\tau}^{i})\times \{0,1\}^2$. 
 
Let $\iota_{t^+}^a = \{x_{1:t}^0,u_{1:t-1}^{a},z_{1:t}^{er},m_{1:t}^{1:2},y_{1:t}\}$ be a realization of the adversary's information $I_{t^+}^a$ (which is the same as the common information at time $t^+$). Define the distribution $\Psi_{t^+}(\iota_{t^+}^a)$ over the space $\prod_{\tau = 1}^t\prod_{i=1,2}\X^i_{\tau} \times\U_{\tau}^{i}$ as follows:
\begin{align*}
\Psi_{t^+}&(\iota_{t^+}^a; x_{1:t},u_{1:t}) \doteq\notag\\
&\Py^{f,g,h^a}[X_{1:t},U_{1:t} = (x_{1:t},u_{1:t}) \mid I_{t^+}^a = \iota_{t^+}^a],
\end{align*}
if $\iota_{t^+}^a$ is \emph{feasible}, that is $\Py^{f,g,h^a}[I_{t^+}^a = \iota_{t^+}^a] > 0$, under the \emph{open-loop} strategy $h^a \doteq (u^a_{1:t-1})$ for the adversary. Otherwise, define $\Psi_{t^+}(\iota_{t^+}^a; x_{1:t},u_{1:t})$ to be the uniform
 distribution over the space $\prod_{\tau = 1}^t\prod_{i=1,2}\X^i_{\tau} \times\U_{\tau}^{i}$.

\begin{lemma}\label{claimJ1}
Let $f,g$ be the team's strategy and let $g^a$ be an arbitrary strategy for the adversary. Then for any realization $x_{1:t},u_{1:t},m_{t}$ of the variables $X_{1:t},U_{1:t},M_t$, we have
\begin{align*}
&\Py^{f,g,g^a}[X_{1:t},U_{1:t-1},M_t = (x_{1:t},u_{1:t-1},m_t) \mid I_t^a] =\notag\\
&\Psi_t(I_t^a; x_{1:t},u_{1:t-1},m_t)\\
&\Py^{f,g,g^a}[X_{1:t},U_{1:t} = (x_{1:t},u_{1:t}) \mid I_{t^+}^a] = \Psi_{t^+}(I_{t^+}^a; x_{1:t},u_{1:t}),
\end{align*}
almost surely.
\end{lemma}
\begin{corollary}\label{factorcorol}
At any given time $t$, the functions $\psi_t$ and $\psi_{t^+}$ can be factorized as
\begin{align}
    \Psi_t(\iota_t^a; x_{1:t},u_{1:t-1},m_t)& = \Psi_t^1(\iota_t^a; x^1_{1:t},u^1_{1:t-1},m^1_t)\notag\\
    &\times \Psi_t^2(\iota_t^a; x^2_{1:t},u^2_{1:t-1},m^2_t) 
    \end{align}
    \begin{align}
    \Psi_{t^+}(\iota_{t^+}^a; x_{1:t},u_{1:t}) &= \Psi^1_{t^+}(\iota_{t^+}^a; x^1_{1:t},u^1_{1:t})\notag\\
    &\times \Psi^2_{t^+}(\iota_{t^+}^a; x^2_{1:t},u^2_{1:t}). 
\end{align}
This is a direct consequence of Lemma \ref{LEM:INDEPEN2}.
\end{corollary}
For any instance $\iota_t^a$ of the adversary's information $I_t^a$, define the distribution $\Phi_t(\iota_t^a)$ over the space $\X_t \times \{0,1\}^2$ as follows
\begin{align}
\Phi_t(\iota_t^a; x_t, m_t) = {\sum_{x_{1:t-1}}\sum_{u_{1:t-1}}\Psi_t(\iota_t^a;x_{1:t},u_{1:t-1},m_t)}.\label{eq:phia}
\end{align}
Similarly, for any instance $\iota_{t^+}^a$ of the adversary's information $I_{t^+}^a$, define the distribution $\Phi_{t^+}(\iota_{t^+}^a)$ over the space $\X_t \times \U_t$ as follows
\begin{align}
\Phi_{t^+}(\iota_{t^+}^a; x_t, u_t) = {\sum_{x_{1:t-1}}\sum_{u_{1:t-1}}\Psi_{t^+}(\iota_{t^+}^a;x_{1:t},u_{1:t})}.\label{eq:phi}
\end{align}
Using Corollary \ref{factorcorol}, we can say that the functions $\Phi_t$ and $\Phi_{t^+}$ can be factorized as
\begin{align}
    \Phi_t(\iota_t^a; x_t, m_t) &= \Phi_t^1(\iota_t^a; x_t^1, m_t^1)\Phi_t^2(\iota_t^a; x_t^2, m_t^2)\\
    \Phi_{t^+}(\iota_{t^+}^a; x_t, u_t) &=  \Phi_{t^+}^1(\iota_{t^+}^a; x_t^1, u_t^1)\Phi_{t^+}^2(\iota_{t^+}^a; x_t^2, u_t^2).
\end{align}

Define communication strategy $\bar{f}^i$ for agent $i$ in the Team such that for any realization $x_t^i, \iota_t^a$ of state $X_t^i$ and adversary's information $I_t^a$ at time $t$, the probability of selecting an action $m_t^i$ at time $t$ is

$\bar{f}_t^i(x_t^i,\iota_{t}^a;m_t^i) \doteq$
\begin{align}\label{eq:defgbara}
\begin{cases}
\frac{\Phi_t^i(\iota_t^a; x_t^i, m_t^i)}{\sum_{m'_t}\Phi_t^i(\iota_t^a; x_t^i, m'_t)} & \text{if } {\sum_{m'_t}\Phi_t^i(\iota_t^a; x_t^i, m'_t)} > 0\\
\mathscr{U}(\cdot) & \text{otherwise},
\end{cases}
\end{align}
where $\mathscr{U}(\cdot)$ denotes the uniform distribution over the action space $\{0,1\}$. Notice that the construction of the strategy $\bar{f}^i$ does not involve adversary's strategy $g^a$.

Define control strategy $\bar{g}^i$ for agent $i$ in the team such that for any realization $x_t^i, \iota_{t^+}^a$ of state $X_t^i$ and adversary's information $I_{t^+}^a$ at time ${t^+}$, the probability of selecting an action $u_t^i$ at time ${t^+}$ is

$\bar{g}_{t}^i(x_t^i,\iota_{{t^+}}^a;u_{t}^i) \doteq $
\begin{align}\label{eq:defgbar}
\begin{cases}
\frac{\Phi_{t^+}^i(\iota_{t^+}^a; x_t^i,u_{t}^i)}{\sum_{u'_{t}}\Phi_{t^+}^i(\iota_{t^+}^a; x_t^i,u'_{t})} & \text{if } \sum_{u'_{t}}\Phi_{t^+}^i(\iota_{t^+}^a; x_t^i,u'_{t}) > 0\\
\mathscr{U}(\cdot) & \text{otherwise},
\end{cases}
\end{align}
where $\mathscr{U}(\cdot)$ denotes the uniform distribution over the action space $\U_t^i$. Notice that the construction of the strategy $\bar{g}^i$ does not involve adversary's strategy $g^a$.

For convenience, let us define
\begin{align}
    \bar{f}_t(x_t,\iota_t^a;m_t) &= \bar{f}_t^1(x_t^1,\iota_t^a;m_t^1)\bar{f}_t^2(x_t^2,\iota_t^a;m_t^2)\\
    \bar{g}_t(x_t,\iota_{t^+}^a;u_t) &= \bar{g}_t^1(x_t^1,\iota_{t^+}^a;u^1_t)\bar{g}_t^2(x_t^2,\iota_{t^+}^a;u_t^2).
\end{align}

\begin{lemma}\label{aseqstrat}
For any strategy $g^a$ for adversary, we have
\begin{align*}
\Py^{((f,g),g^a)}[M_t = m_t \mid X_t, I_t^a] = \bar{f}_t(X_t,I_{t}^a;m_t)
\end{align*}
\begin{align*}
\Py^{((f,g),g^a)}[U_t = u_t \mid X_t, I_{t^+}^a] = \bar{g}_{t^+}(X_t,I_{{t^+}}^a;u_t)
\end{align*}
almost surely for every $m_t,u_t$.
\end{lemma}
\begin{proof}
Let $x_t, \iota_t^a, \iota_{t+}^a$ be a realization that has a non-zero probability of occurrence under the strategy profile $((f,g),g^a)$. Then using Lemma \ref{claimJ1}, we have
\begin{align}
\Py^{((f,g),g^a)}&[X_{1:t},U_{1:t-1},M_t = (x_{1:t},u_{1:t-1},m_t) \mid \iota_t^a] \notag\\
&= \Psi_t(\iota_t^a; x_{1:t},u_{1:t-1},m_t), \label{eq:lemma14a}
\end{align}
\begin{align}
\Py^{((f,g),g^a)}&[X_{1:t},U_{1:t} = (x_{1:t},u_{1:t}) \mid \iota_{t+}^a] \notag\\
&= \Psi_{t+}(\iota_{t+}^a; x_{1:t},u_{1:t}), \label{eq:lemma14}
\end{align}
for every realization $x_{1:t-1}$ of states $X_{1:t-1}$ and $u_{1:t-1}$ of action $U_{1:t-1}$. Summing over all $x_{1:t-1},u_{t}$ and using \eqref{eq:phi}, \eqref{eq:phia}, \eqref{eq:lemma14a} and \eqref{eq:lemma14},  we have
\begin{align}
 \Py^{((f,g),g^a)}[X_t = x_t \mid I_t^a = \iota_t^a] =\sum_{m_t}\Phi_t(\iota_t^a; x_t,m_t).
\end{align}
\begin{align}
 \Py^{((f,g),g^a)}[X_t = x_t \mid I_{t+}^a = \iota_{t+}^a] =\sum_{u_t}\Phi_{t+}(\iota_{t+}^a; x_t,u_t).
\end{align}
The left hand side of the above equation is positive since $x_t,i^a_t, \iota_{t+}^a$ is a realization of positive probability under the strategy profile $((f,g),g^a)$.

Using Bayes' rule,  \eqref{eq:phia}, \eqref{eq:defgbara} and \eqref{eq:lemma14a}, we obtain
\begin{align}
\nonumber\Py^{((f,g),g^a)}[M_t = m_t \mid X_t = x_t, I_t^a = \iota_t^a]
&= \bar{f}_t(x_t,\iota_{t}^a;m_t) .
\end{align}
Using Bayes' rule,  \eqref{eq:phi}, \eqref{eq:defgbar} and \eqref{eq:lemma14}, we obtain
\begin{align}
\nonumber\Py^{((f,g),g^a)}[U_t = u_t \mid X_t = x_t, I_{t+}^a = \iota_{t+}^a]
&= \bar{g}_t(x_t,\iota_{t+}^a;u_t) .
\end{align}
This concludes the proof of the lemma.
\end{proof}

Let us define $((\bar{f},\bar{g}),g^a)$, where $\bar{f},\bar{g}$ is as defined in \eqref{eq:defgbara}, \eqref{eq:defgbar}. We can now show that the strategy $(\bar{f},\bar{g})$ satisfies
\[
J((\bar{f},\bar{g}),g^a) = J(({f},{g}),g^a),
\]
for every strategy $g^a \in \mathcal{G}^a$. 
Because of the structure of the cost function in \eqref{eq:cost2}, it is sufficient to show that for each time $t$, the random variables   $(X_t,M_t,I^a_t)$ have the same joint distribution under strategy profiles $(({f},{g}),g^a)$ and $((\bar{f},\bar{g}),g^a)$ and at time $t^+$, the random variables $(X_t,U_t,U_t^a,I^a_{t^+})$ have the same joint distribution under strategy profiles $(({f},{g}),g^a)$ and $((\bar{f},\bar{g}),g^a)$. We prove this by induction. It is easy to verify that at time $t=1$, $(X_1,M_1,I^a_1)$ have the same joint distribution under strategy profiles $(({f},{g}),g^a)$ and $((\bar{f},\bar{g}),g^a)$.
Now assume that at time $t$, 
\begin{align}
\label{jointeq}\Py^{(({f},{g}),g^a)}[x_t,m_t,\iota_t^a] = \Py^{((\bar{f},\bar{g}),g^a}[x_t,m_t,\iota_t^a],
\end{align}
for any realization of state, actions and adversary's information $x_t,m_t,\iota_t^a$. Let $\iota_{t^+}^a = (\iota_{t}^a,z^{er}_t,m_{t},y_t)$. Then we have
\begin{align}
&\Py^{(({f},{g}),g^a)}[x_{t},\iota_{t^+}^a] =\notag\\
&\Py[y_{t} \mid z^{er}_{t},{x}_t,\iota_{t}^a,m_t]\Py[z^{er}_{t} \mid {x}_t,\iota_{t}^a,m_t]\Py^{(({f},{g}),g^a)}[{x}_t,\iota_{t}^a,m_t]\\
&= \label{indhyp}\Py[y_{t} \mid z^{er}_{t},{x}_t,\iota_{t}^a,m_t]\Py[z^{er}_{t} \mid {x}_t,\iota_{t}^a,m_t]\notag\\
&\times\Py^{((\bar{f},\bar{g}),g^a}[{x}_t,\iota_{t}^a,m_t]\\
&=\Py^{((\bar{f},\bar{g}),g^a}[x_{t},\iota_{t^+}^a].\label{indhyp2}
\end{align}
At ${t}^+$, for any realization $x_{t},u_{t},u_{t}^a,\iota_{{t}^+}^a$ that has non-zero probability of occurrence under the strategy profile ${(({f},{g}),g^a)}$, we have
\begin{align}
\label{constarg}&\Py^{(({f},{g}),g^a)}[x_{t},u_{t},u_{t}^a,\iota_{{t}^+}^a]=\Py^{(({f},{g}),g^a)}[x_{t},\iota_{{t}^+}^a]g_t^a(\iota_{{t}^+}^a;u_{t}^a) \notag\\
&\times\Py^{(({f},{g}),g^a)}[u^1_{t}\mid x^1_{t},\iota_{{t}^+}^a]\Py^{(({f},{g}),g^a)}[u^2_{t}\mid x^2_{t},\iota_{{t}^+}^a]\\
&= \Py^{(({f},{g}),g^a)}[x_{t},\iota_{{t}^+}^a]g_t^a(\iota_{{t}^+}^a;u_{t}^a)\bar{g}_t^{1}(x^1_{t},\iota_{{t}^+}^a;u_{t}^{1})\bar{g}_t^{2}(x^2_{t},\iota_{{t}^+}^a;u_{t}^{2})\label{constarg1}\\
&= \Py^{((\bar{f},\bar{g}),g^a}[x_{t},\iota_{{t}^+}^a]g_t^a(\iota_{{t}^+}^a;u_{t}^a)\bar{g}_t^{1}(x^1_{t},\iota_{{t}^+}^a;u_{t}^{1})\bar{g}_t^{2}(x^2_{t},\iota_{{t}^+}^a;u_{t}^{2})\label{constarg3}\\
&= \Py^{((\bar{f},\bar{g}),g^a}[x_{t},\iota_{{t}^+}^a]g_t^a(\iota_{{t}^+}^a;u_{t}^a)\Py^{((\bar{f},\bar{g}),g^a)}[u^1_{t}\mid x^1_{t},\iota_{{t}^+}^a]\notag\\
&\times\Py^{((\bar{f},\bar{g}),g^a)}[u^2_{t}\mid x^2_{t},\iota_{{t}^+}^a]\label{constarg2}\\
&= \Py^{((\bar{f},\bar{g}),g^a}[x_{t},u_{t},u_{t}^a,\iota_{{t}^+}^a]\label{constarg4},
\end{align}
where the equality in \eqref{constarg} is a consequence of the chain rule and the manner in which players randomize their actions. Equality in \eqref{constarg1} follows from Lemma \ref{aseqstrat} and the equality in \eqref{constarg3} follows from the result in \eqref{indhyp2}.
 Then we have
\begin{align}
&\Py^{(({f},{g}),g^a)}[x_{t+1},\iota_{t+1}^a] =\notag\\
&\nonumber\sum_{\bar{x}_t}\sum_{\bar{u}_t}\Py[x_{t+1},x_{t+1}^0 \mid \bar{x}_t,\bar{u}_t,u_t^{a},\iota_{t^+}^a]\Py^{(({f},{g}),g^a)}[\bar{x}_t,\bar{u}_t,u_t^{a},\iota_{t^+}^a]\\
&= \label{indhyp}\sum_{\bar{x}_t}\sum_{\bar{u}_t}\Py[x_{t+1},x_{t+1}^0 \mid \bar{x}_t,\bar{u}_t,u_t^{a},\iota_{t^+}^a]\notag\\&\quad \quad \times\Py^{((\bar{f},\bar{g}),g^a}[\bar{x}_t,\bar{u}_t,u_t^{a},\iota_{t^+}^a]\\
&=\Py^{((\bar{f},\bar{g}),g^a}[x_{t+1},\iota_{t+1}^a].\label{indhyp2}
\end{align}
The equality in (\ref{indhyp}) is due to the induction hypothesis. Note that the conditional distribution $\Py[x_{t+1},x_{t+1}^0 \mid \bar{x}_t,{\bar{u}_t},u_t^{a},\iota_{t^+}^a]$ does not depend on players' strategies.
At $t+1$, for any realization $x_{t+1},m_{t+1},\iota_{t+1}^a$ that has non-zero probability of occurrence under the strategy profile ${(({f},{g}),g^a)}$, we have
\begin{align}
\label{consarg}&\Py^{(({f},{g}),g^a)}[x_{t+1},m_{t+1},\iota_{t+1}^a] \\
&= \Py^{(({f},{g}),g^a)}[m_{t+1}\mid x_{t+1},\iota_{t+1}^a]\Py^{(({f},{g}),g^a)}[x_{t+1},\iota_{t+1}^a]\\
&= \bar{f}_t(x_{t+1},\iota_{t+1}^a;m_{t+1})\Py^{((\bar{f},\bar{g}),g^a)}[x_{t+1},\iota_{t+1}^a]\label{consarg1}\\
&= \Py^{((\bar{f},\bar{g}),g^a)}[m_{t+1}\mid x_{t+1},\iota_{t+1}^a]\Py^{((\bar{f},\bar{g}),g^a)}[x_{t+1},\iota_{t+1}^a]\label{consarg3}\\
&= \Py^{((\bar{f},\bar{g}),g^a}[x_{t+1},m_{t+1},\iota_{t+1}^a]\label{consarg4},
\end{align}

Therefore, by induction, the equality in \eqref{jointeq} holds for all $t$. This concludes the proof of Claim \ref{claim:one}. \qed

\section{Proof of Lemma \ref{beliefprop}}\label{beliefpropproof}
Let us define the team's common information belief on its agents' private states as
\begin{align*}
    B_t(x^0,x^1,x^2) &= \prob(X_t^0 = x_t^0,X^1_t=x^1, X^2_t=x^2|\cteam_t).\\
    B_{t^+}(x^0,x^1,x^2) &= \prob(X_t^0 = x_t^0,X^1_t=x^1, X^2_t=x^2|\cteam_{t^+}).
\end{align*}
Note that this belief coincides with the common information belief when the adversary has maximum information. Therefore, we can use the transformations $\bar{\eta}_t$ and $\bar{\beta}_t$ (see Section \ref{commencryptsec}) to update the belief $B_t$. 

\begin{lemma}\label{belprivatered}
There exists a min-max strategy for the team of the form
\begin{align}
     M^i_t &\sim  f_t^i(X_t^i,B_t,C_t)\\
      U^i_t &\sim {g}^i_t(X^i_{t},B_{t^+},C_{t^+})\label{teambeliefstrat}
\end{align}
\end{lemma}
\begin{proof}
This lemma can be shown by replacing the agents in the team with a virtual coordinator \cite{nayyar2013decentralized}. The coordinator only sees the team's common information and selects prescriptions for both agents which will be used by the agents to map their private information $X_t^i$ to an action. When the agents in the team are replaced with a virtual coordinator, the team-game can be viewed as a zero-sum game between two individual players: the coordinator and the adversary. It was shown in \cite{gensbittel2014existence} that the more-informed player (coordinator) can select its actions (prescriptions) based on its belief on the system state ($B_t$) and the adversary's information ($C_t$) without loss of optimality.
\end{proof}
Thus, the private information used by the agent $i$ in the team has been further reduced to $X_t^i,B_t$ from $X_t^i, D_t$.

\begin{lemma}\label{beliefprop}
Let $(f,g)$ be a belief based strategy as in \eqref{teambeliefstrat} for the team. For any such strategy pair, the team's belief $B_t$ is given by
\begin{align}
    B_t &= \varpi_t(X_{L_t},C_t)\label{teambel1}\\
    B_{t^+} &= \varpi_{t^+}(X_{L_{t^+}},C_{t^+})\label{teambel2},
\end{align}
where $\varpi_t$ and $\varpi_{t^+}$ are transformations that may depend on the team's strategy $f,g$.
\end{lemma}
\begin{proof}
We first split the team's strategy into the coordinator's strategy and prescriptions in the following manner. The prescriptions for agent $i$ are defined as
\begin{align}
    \Gamma_t^i &= f_t^i(\cdot, B_t,C_t)\\
    \Lambda_t^i &= g_t^i(\cdot, B_{t^+},C_{t^+}).
\end{align}
The pair of prescriptions is denoted by $\Gamma_t := \vartheta_t(B_t,C_t)$ and $\Lambda_t= \vartheta_{t^+}(B_{t^+},C_{t^+})$. Here, $\vartheta$ represents the coordinator's strategy.

We prove this lemma by induction. Our induction hypothesis is that \eqref{teambel1} holds at time $t$. Note that at $t=1$, our induction hypothesis holds because the team's common information $\cteam_1$ and the overall common information $C_1$ are identical. We now prove that \eqref{teambel2} holds at time $t^+$. The new common observations received by the team at time $t^+$ are $(Z_{t^+},Z_t^{er})$, where $Z_{t^+}$ denotes the adversary's observations at time $t^+$. If $Z_t^{er} \neq \varnothing$, then $Z_t^{er} = X_t = X_{L_{t^+}}$. Therefore, $B_{t^+}$ is simply the degenerate distribution, i.e. $B_{t^+}(x) = \mathds{1}_{X_{t}}(x)$ Thus, \eqref{teambel2} holds. If $Z_t^{er} = \varnothing$, we have
\begin{align}
    B_{t^+} &\stackrel{a}{=} \bar{\eta}_{t}(B_t,\Gamma_t,(Z_{t^+},\varnothing))\\
    &= \bar{\eta}_{t}(B_t,\vartheta_t(B_t,C_t),(Z_{t^+},\varnothing))\\
    &\stackrel{b}{=:}\varpi_{t^+}(X_{L_t},C_{t^+})\\
    &=\varpi_{t^+}(X_{L_t^+},C_{t^+}).\label{indplus}
\end{align}
Here, $(a)$ follows from the fact that we can use $\bar{\eta}_t$ in Section \ref{commencryptsec} to update $B_t$.
Equality in $(b)$ is because by induction hypothesis at time $t$, $B_t=\varpi_t(X_{L_t},C_t)$.

We now prove that the induction hypothesis holds at time $t+1$. The common observations at time $t+1$ are $Z_{t+1}$. Using the belief update transformation $\bar{\beta}_t$ in Section \ref{commencryptsec}, 
\begin{align}
    B_{t+1} &= \bar{\beta}_t(B_{t^+},\Lambda_t,Z_{t+1})\\
    &= \bar{\beta}_t(B_{t^+},\vartheta_{t^+}(B_{t^+},C_{t^+}),Z_{t+1})\\
    &\stackrel{c}{=:} \varpi_{t+1}(X_{L_{t^+}},C_{t+1})\\
    &=\varpi_{t+1}(X_{L_{t+1}},C_{t+1}).
\end{align}
Here, $(c)$ is a consequence of \eqref{indplus}.
Therefore, by induction, the lemma holds at all times $t$ and $t^+$.
\end{proof}

Proposition \ref{privateredencrypt} can be obtained by substituting $B_t$ in Lemma \ref{belprivatered} with the simplified expressions in Lemma \ref{beliefprop}.

\end{document}